# DSBplot: Indels in DNA Double-strand Break Repair Experiments

**Tejasvi Channagiri[1,2], Margherita Maria Ferrari[1,3,6], Youngkyu Jeon[4,5,6], Penghao Xu[4,6], Francesca Storici[4], Nataša Jonoska[1]**

[1]Department of Mathematics and Statistics, University of South Florida, Tampa, Florida; [2]currently at Department of Biostatistics, Harvard University; [3]currently at Department of Mathematics, University of Manitoba, Winnipeg, Canada; [4]School of Biological Sciences, Georgia Institute of Technology, Atlanta, Georgia; [5]Molecular Targets Program, National Cancer Institute, National Institutes of Health, Frederick, MD; [6]these authors contributed equally.

**Abstract**

Double-strand breaks (DSBs) in DNA are naturally occurring destructive events in all organisms that may lead to genome instability. Cells employ various repair methods known as non-homologous end joining (NHEJ), microhomology mediated end joining (MMEJ), and homology-directed recombination (HDR). These repair processes may lead to DNA sequence variations (e.g., nucleotide insertions, deletions, and substitutions) at the location of the break. Studying DNA DSB repair processes often involves the use of high throughput sequencing assays to precisely quantify the sequence variations near the break with software tools. Often methods of assessing and visualizing these data have not taken into account the full complexity of the sequencing data, such as the frequency, type, and position of the sequence variations in a single comprehensive representation. Here we present a method that allows visualization of the overall variation pattern as well as comparison of these patterns among experimental setups.

**Introduction**

In recent years, DNA repair of double-strand break (DSB) and genome editing promoted by DSBs have been studied by DNA high throughput sequencing of the genetic material obtained after the DNA repair of experimentally induced breaks, frequently by CRISPR/Cas9 assays, at specific locations in a genetic sequence[1–3]. In the case of breaks induced in DNA within cell environments, the repair may be done by one of the natural repair processes, including the error prone *non-homologous end-joining* (NHEJ) or *microhomology mediated end-joining* (MMEJ) pathways, or the precise homology-directed repair (HDR) pathway that could be used for genome editing[4–6]. To study the repaired DNA, a process known as *targeted amplicon sequencing* (or *targeted next generation sequencing* (*NGS*), *amplicon sequencing*, *amplicon deep sequencing*, etc.) is used, where primers are applied near locations around the DSB that allow amplification of the sequence around the repaired DSB that is then sequenced with high-throughput sequencing technologies[7]. The resulting sequencing reads are aligned to the expected repair outcome, called the *reference sequence* (or *reference amplicon*, *amplicon sequence*, *amplicon*, etc.), to detect and quantify the variations (insertions, deletions, and substitutions) near the DSB site resulting from the repair process. The huge amount of data produced by high-throughput sequencing platforms, such as Illumina HiSeq[8], call for automated tools and various software to analyze the data. Such software tools include `CRISPR-GA`[9], `AGEseq`[10], `CrispRVariants`[11], `CRISPResso`[12], `BATCH-GE`[13], `Cas-Analyzer`[14], `CRISPR-DAV`[15], `Hi-FiBR`[16], `RIMA`[17], `CRISPResso2`[18], `ampliCan`[19], and `PEM-Q`[20]. However, these tools have not considered the full information contained in the sequencing data and often summarize the distribution of insertions and deletions (indels) using figures focused on a couple of properties of the data. For example, such figures may include 2D histograms showing indel frequency vs. positions on the reference sequence or indel frequency vs. length of indel (e.g., `CRISPR-GA`, `CRISPResso`, `Cas-Analyzer`, `CRISPR-DAV`, and `CRISPResso2`) though some tools also show a list of the most frequently occurring variants aligned to the reference sequence (e.g., `CrispRVariants`, `BATCH-GE`, `Cas-Analyzer`, `RIMA`, `CRISPResso2`, and `ampliCan`).

Here we present `DSBplot`, a software tool developed to represent sequence variations and their respective frequencies in libraries obtained from targeted amplicon sequencing. Our method expands on the histogram or table summaries by allowing a detailed visualization of the indels and substitutions, including their frequency, in the sequencing data. This method was applied in Jeon *et al.*[21] to study the effect of an endogenous RNA transcript on DSB DNA repair in human cells. The study involved the analysis of dozens of sequencing libraries, each of which contained several million reads. To quickly quantify the indels in the sequencing data, an automated pipeline aligned the raw reads against a reference sequence and produced visual summaries to guide further analyses. The method showed the differential abundance of rarely occurring variants alongside the commonly occurring ones, therefore capturing even small differences in the indel distributions between the experimental constructs. The resulting figures proved to be invaluable as they revealed salient patterns in the data such as prevalent single nucleotide insertions in some assays possibly due to a sticky-end cut after cleavage, or the impact of the location of the DSB and its sequence context on the indel distribution.

**`DSBplot` Protocol**

The general steps included in `DSBplot` are depicted in the flowchart in **Figure 1**. We describe them in more detail in the following.

*Input*

Users have two options for the input to `DSBplot`. They may either provide FASTQ files representing sequencing reads with quality information or they may provide pre-aligned SAM[22] (Sequence Alignment/Map) files. In both cases they must also supply a single reference sequence in FASTA format (more than one reference sequence is not allowed). When supplying FASTQ files as input, it is expected that the user has already trimmed and quality filtered the reads. When supplying SAM files as input, it is expected that the user has already aligned the reads against the reference sequence and that the reference sequence provided as input to the command is identical to the one used for creating the SAM file. In either case, it is assumed that each FASTQ or SAM file corresponds to an independent replicate of the same experiment. In the final output, the frequencies in the replicates are averaged, as described later. It is also assumed that the reference sequence is chosen to correspond to the reads in such a way that the 5′ end of the reads align exactly with the 5′ end of the reference sequence. Due to sequencing read length limitations, reads may be shorter than the reference sequence, so the 3′ ends do not have to match exactly. However, the alignment with the reference sequence must extend about 30 nucleotides past the DSB site without too many indel/substitutions to pass the filtering stage, as discussed later.

*Processing*

The processing portion of the tool, run with the command `DSBplot-process`, is similar to other tools that study products of CRISPR/Cas9 assays in the context of genome editing. If FASTQ files are used as input, the processing starts by aligning the raw reads against the reference sequence. The method built into the DSBplot pipeline uses the popular alignment software `Bowtie 2`[23] (https://bowtie-bio.sourceforge.net/bowtie2/manual.shtml) to produce aligned SAM files. If pre-aligned SAM files are used as input instead, then the alignment with `Bowtie 2` is omitted. After obtaining the SAM files, custom scripts are used to realign the sequences to try to ensure that the indels are consecutive and overlap the DSB site (i.e., when there are multiple, equally good alignments, we try to select the one that satisfy these criteria). The NHEJ pathway is expected to result in variations near the DSB repair site[6], and thus an alignment with variations closer to the DSB site should be more biologically suitable. Other optional filtering criteria are also possible, such as setting the maximum number of allowed substitutions in an alignment and/or the minimum length of a read. Our experience has shown that too many substitutions or too short of a read may indicate that the alignment is faulty (e.g., the read may come from extraneous DNA). The filtering and realignment stages output detailed tables indicating reads that were accepted or rejected, as well as a summary table of the number of reads accepted/rejected due to each filtering criteria. This allows users to debug possible alignment and processing problems or readjust the filtering. There are several optional command-line parameters to allow omitting or modifying each part of the filtering, such as the consecutive indel criteria, indel overlapping DSB criteria, and whether or not to perform realignment (see README at https://github.com/tchannagiri/DSBplot#readme).

To focus on indels around the DSB site that are characteristic of NHEJ repair and ignore large variations at the ends of reads, we do the following. In each read we concentrate on a sequence window around the DSB site, called the *repair window*. By default, the repair window consists of the nucleotides on the read that align with the 10 nucleotides upstream and 10 nucleotides downstream of the DSB site on the reference sequence. The number of nucleotides in the repair window may vary in each read depending on the number of indels in the alignment (e.g., insertions will increase the number of nucleotides and deletions will decrease the number of nucleotides). Optionally, to ensure that the indels do not extend past the window, we may check two *anchor sequences* on each side of the repair window. These are the nucleotides on the read that align to the 20 nucleotides upstream of the repair window and the 20 nucleotides downstream of the repair window on the reference sequence. **Figure 2A** shows an example read aligned to the reference sequence, and a demarcation of the repair window and anchor sequences. Since we expect that most DSB repair studies will focus on the repair windows, expecting minimal variations away from the DSB site, the default `DSBplot` setting requires a match of each anchor sequence with at most one substitution and no indels. Therefore, the alignment shown in **Figure 2A** will be rejected in the default setting because there is an insertion in the left anchor sequence.

The frequencies of the reads with the same repair window are then summed and a table containing each unique repair window with its corresponding frequency in the given experiment is outputted in a repair window table. The process of obtaining repair windows and checking anchor sequences is highly customizable and can be modified or omitted by specifying the appropriate command-line arguments (e.g., for the repair window size, anchor sequence size, permissible number of indels/substitutions on the anchor sequences, etc.). If multiple independent experimental replicates were specified, the frequencies of their repair windows are averaged in the output table. In our experience, alignment substitutions were an unreliable marker to study patterns of DSB DNA repairs since they occur systematically as sequencing errors[24–27], and so two versions of the repair window tables are created, one with and another without substitutions. In the without-substitution table, the detected substitutions in the repair windows are replaced by the corresponding nucleotides on the reference sequence, while the with-substitution table contains the subsitutions as is. The process of obtaining repair windows around the DSB facilitates the study of variations due to NHEJ, which usually results in small variations near the DSB[6], but larger variations, such as large deletions due to MMEJ[5], may be filtered out by `DSBplot` and will require a separate analysis.

After the repair windows are obtained, a *repair variation* table showing individual variations for each repair window is created. Each repair window is first separated into its constituent variations as shown in **Figure 2C**, where a single repair window is separated into seven individual variations. Though the figure shows only a single repair window, this process is carried out for all repair windows in the repair window table. Each individual variation is associated with the following data:

- Position: The position relative to the DSB site.
- Number of variations: The number of variations on the repair window that the variation originated from. This will be the same for all variations originating from the same repair window as shown in **Figure 2C**.
- Type: The type of variation (insertion, deletion, or substitution).
- Frequency: The frequency of the repair window that the variation originated from. This will be the same for all variations originating from the same repair window as shown in **Figure 2C**.

Once all variations from all repair windows are tabulated, the variations are grouped by the position, number of variations, and type, and the frequencies are summed to obtain the final repair variation table. Like the repair window table, the repair variation table is outputted in two versions, with and without substitutions. These tables are used to create *variation-position histograms*, described later.

The `DSBplot-process` also creates a metadata table in the output directory showing the original reference sequence used for alignment, the reference sequence when restricted to the specified window, the DSB position on the reference sequence, and other fields that may be needed for plotting the variation-distance graphs and variation-position histograms.

*Variation-distance graphs and variation-position histograms*

After `DSBplot-process` has output the tables, the commands `DSBplot-graph` and `DSBplot-histogram` can be applied to these tables. These commands organize the data into *variation-distance graphs* and *variation-position histograms*, respectively. In variation-distance graphs the vertices represent the repair windows detected in the data, while two repair windows that differ by a single indel or substitution are connected by an edge (**Figure 2B**). The graph formulation allows, for example, using a layout algorithm that clusters repair windows together depending on the length of the path between them (see the Kamada-Kawaii[28] layout description at https://github.com/tchannagiri/DSBplot#layouts). The graph plotting script contains several options for customizing the output image including the layout of vertices, how the vertices are colored, whether to display legends, the file output type, and several others.

If a single sample is passed to the `DSBplot-graph` command, an *individual* graph is plotted. In an individual graph, the vertex colors indicate the type of variation (insertion, deletion, substitution, mixed, or no variation) and the size of the vertex indicates the log-frequency of the corresponding repair window. If a pair of samples are passed to the `DSBplot-graph` command (separated by "::" on the command line), a *comparison* graph is plotted. In a comparison graph, the vertices are colored using the log-frequency-ratio of the corresponding repair windows (i.e., log(frequency in sample 1 / frequency in sample 2)) for the color gradient. This allows easy comparison of the repair window distribution between two samples. The size of the vertices in a comparison graph reflects the log of the maximum of the two frequencies in the two samples (i.e., log(max(frequency in sample 1, frequency in sample 2))). See **Figure 3** for examples of individual/comparison variation-distance graphs and variation-position histograms. A useful feature of the variation-distance graph script is the ability to output figures in HTML format, which allows users to interactively zoom and pan figures, as well as hover over vertices and edges to see the exact sequence represented by the vertex or connected by the edges.

The variation-position histograms are 3D plots that show the distribution of a given type of variation (insertion, deletion, or substitutions) within each repair window, grouping these variations according to their position on the reference sequence and the number of variations in their respective repair window, and summing their frequencies. In particular, the $x$-coordinate indicates the relative position of the variation relative to the DSB site on the reference sequence, the $y$-coordinates indicates the total number of variations on the repair window (that the variation originated from), and the $z$-axis shows the frequencies of all repair windows with such variations (see **Figure 3B**). For example, a bar of height $z$ in location $(x, y)$ in a variation-position histogram for deletions indicates that all repair windows with $y$ total variations that have a deletion at position $x$, appear with total frequency $z$.

See the included README (https://github.com/tchannagiri/DSBplot#readme) for detailed information about the different visualization and processing options.

**Conclusion**

We presented a pipeline for complete processing of high-throughput sequencing data to study DNA repair variations due to DSBs. While this tool was initially developed to study the DNA DSB repair process itself, the functionality expands on the functionality of similar tools meant for studying genome editing with CRISPR/Cas9 and other systems. In contrast to previous tools that lose much of the richness of sequencing information by simplifying the indel distribution in a single 2D histogram or identifying only the top variants, this tool allows practically all DSB repair events within the chosen window of the DSB site in an amplicon sequencing library to be represented in a single figure. Additionally, the tool allows two experimental constructs to be easily compared. The presentation of the repair window distribution as a graph provided by `DSBplot` opens the door for new types of analyses to be performed on the output. These include various graph analyses, graph clustering tools, as well as approaches arising from topological data analysis.

`DSBplot` has advantages over other similar tools with the flexibility built into the tool through many command-line options, the ability to easily write scripts for batch analysis of many libraries, and the rich output in the form of detailed tables and figures. However, `DSBplot` does have the drawback of being intended for users with command line scripting experience, unlike other tools such as `CRISPR-GA`, `CRISPResso`, `Cas-Analyzer`, and `CRISPResso2`, which have web interfaces. A detailed README of both the protocol and usage can be found in the repository along with demonstration data and scripts. `DSBplot` is built with Python 3 and makes significant use of the packages `NetworkX`[29] (graph algorithms) and `Plotly`[30] (plotting).

**Data and Code Availability**

The data and used to produce the example figures in this article, as well as the code for `DSBplot` is available in the following locations: https://knot.math.usf.edu/software/DSBplot/DSBplot-0.1.3.tar.gz, https://zenodo.org/records/10494583, and https://github.com/tchannagiri/DSBplot. Please see the instructions in the README (https://github.com/tchannagiri/DSBplot#demonstration) for reproducing the figures in this article.

**Acknowledgements**

This research was under auspices of the Southeast Center for Mathematics and Biology, an NSF-Simons Research Center for Mathematics of Complex Biological Systems (NSF grant DMS-1764406 and the Simons Foundation grant 594594 to F.S. and N.J.), as well as the W.M. Keck Foundation. N.J. was partially supported by the NSF grants DMS-2054321, CCF-2107267, and the Simon's Fellow award 917563 from the Simons Foundation. We acknowledge funding from the National Institute of General Medical Sciences (NIGMS) of the NIH (grant GM115927 to F.S.), the National Science Foundation fund (grant MCB-1615335 to F.S.), and the Howard Hughes Medical Institute Faculty Scholar (grant 55108574 to F.S.).

**Author Contributions**

T.C. with help from M.M.F., Y.J., and P.X. wrote the software. T.C. with help from all authors wrote the manuscript. F.S. together with Y.J., M.M.F., and N.J. conceived the original project and designed experiments that led to the development of the software. All authors contributed to the design of the protocol. All authors commented on and approved the manuscript.

**Competing Interests**

The authors declare no competing interests.

**Figure 1**

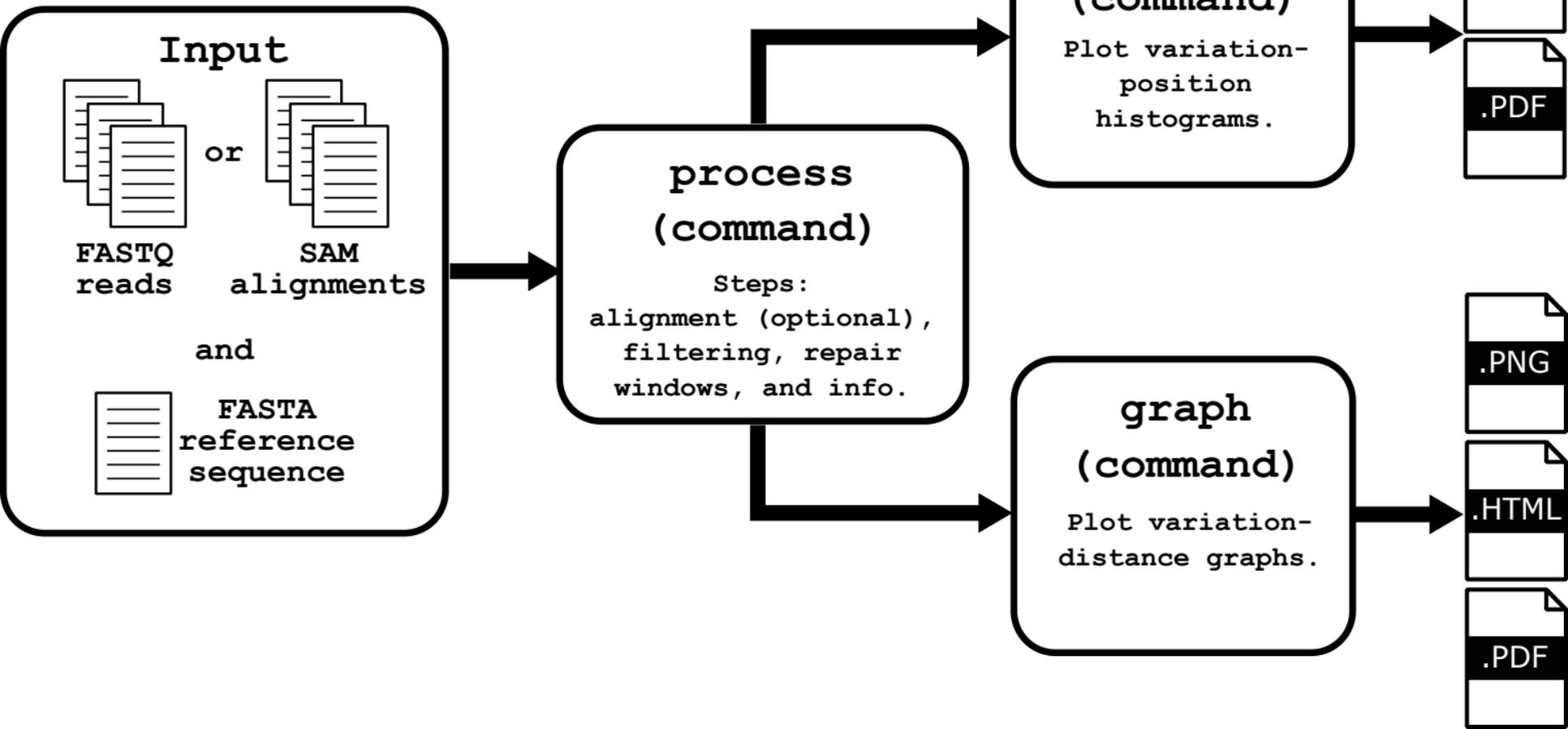

**Figure 1.** Flowchart representing the sequence of steps for using the `DSBplot` commands. The `DSBplot-process` command must first be run with the targeted amplicon sequencing reads in FASTQ format or the aligned reads in SAM format, and the corresponding reference sequence in FASTA format as input. The `DSBplot-process` command generates a directory of tabular data used for plotting the variation-distance graphs and variation-position histograms. The `DSBplot-graph` and `DSBplot-histogram` commands can then be used with this directory as input to produce the variation-distance graphs and variation-position histograms, respectively. The resulting figures may be output in several formats including PNG, PDF, and HTML (for graphs).

**Figure 2**

**A. Repair window extraction**

**B. Graph adjacency relation**

**C. Splitting repair window into variations**

**Figure 2**. **A**, A reference sequence aligned to a read sequence with the repair window region and anchor sequence regions demarcated. In this example, the repair window are the nucleotides of the read sequence that align to the 20 nucleotides around the DSB (from -10 to 10) on the reference sequence. The read sequence has four deletions (from -4 to -1) and two insertions (just right of -1) in the repair window, making the repair window 18 nucleotides in length. The two anchor sequences are the nucleotides on the read that align to the 20 nucleotides upstream of the repair window (from -30 to -11) and the 20 nucleotides downstream (from 11 to 30) of the repair window on the reference sequence, respectively. In this case, the left anchor has one insertion (just right of -21) and the right anchor has one substitution (at 18). **B**, The graph adjacency relationship holds between vertices whose repair windows differ by a single indel or substitution. This is an example path between the vertex with no variations (GACTCCTCCCGACGGCTGCT) and a vertex with seven variations (GACTCCTGACGGGTGCT). The white fill with green outline indicates the vertex with no variations, the blue fill indicates vertices with only deletions, and the green fill indicates vertices with a mix of indels/substitutions. The red arrows/letters indicate the position of the variation. **C**, An example of how a single repair window is split into its constituent variations for creating the variation-position histograms.

# Figure 3

**A**

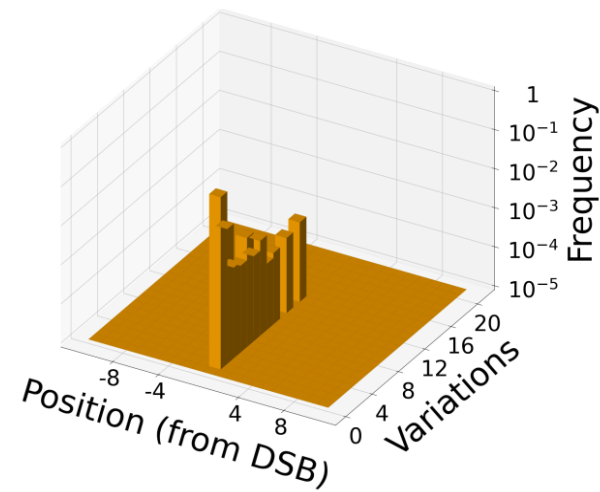
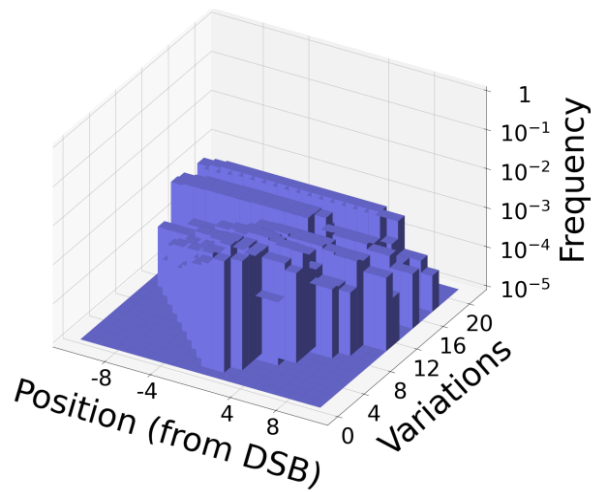
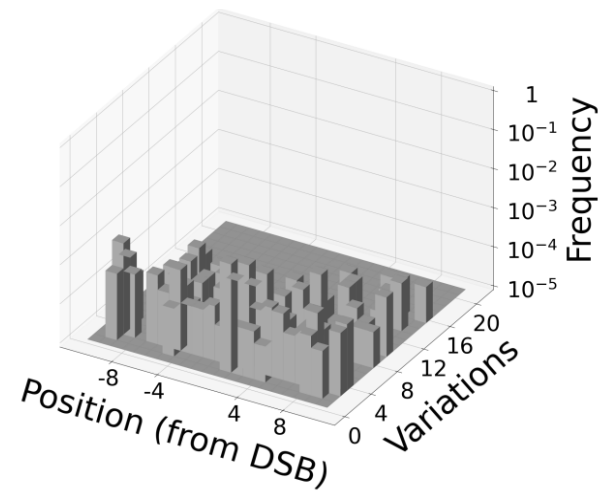

**B**

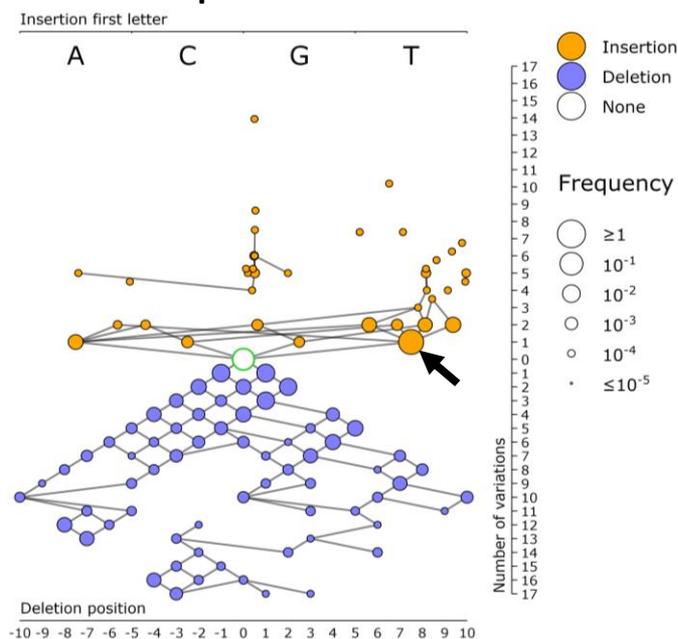
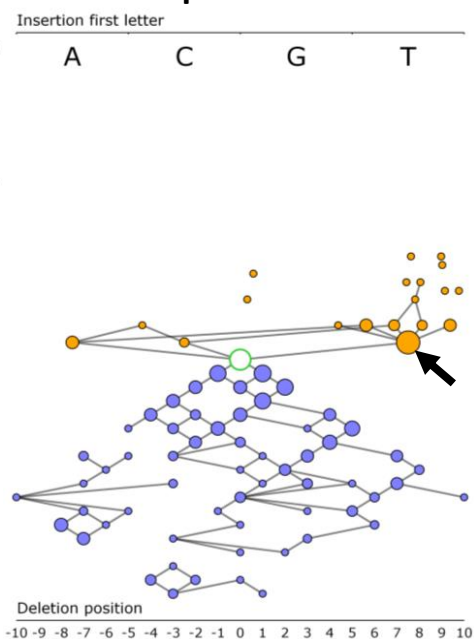
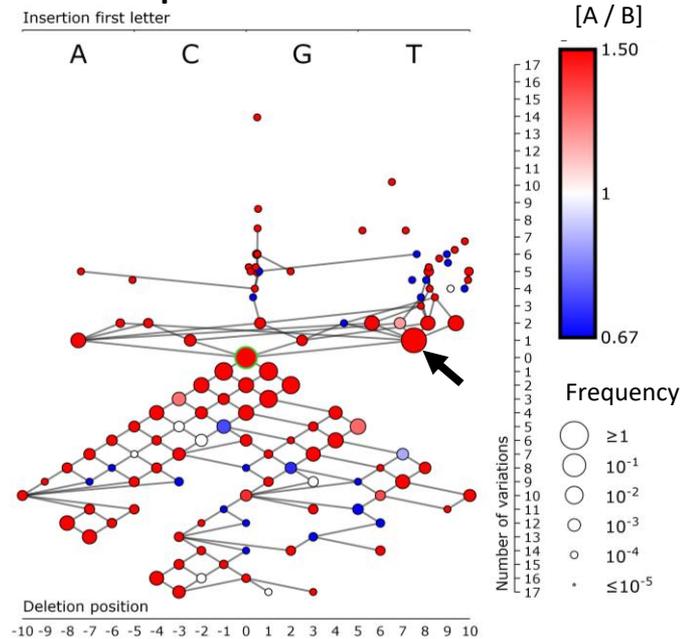

**Figure 3**. **A**, Variation-position graphs for insertions, deletions, and substitutions in Experiment A. The figures are generated with a subset of real data from the Jeon *et al.* study for illustration purposes[21]. For more information about the graph layouts and a demonstration of how to generate these figures, please see the README in the GitHub repository (https://github.com/tchannagiri/DSBplot). **B**, Individual variation-distance graph for Experiment A, individual variation-distance graph for Experiment B, and comparison variation-distance graph showing the ratio of frequencies of Experiment A to Experiment B. In both the individual graphs we note that a single T insertion (black arrow) appears relatively frequently in both experiments but with higher frequency in Experiment A, as the comparison graph reveals (i.e., the vertex is red in the comparison graph). Visually comparing the individual graphs also reveals that there are repair windows that appear in one experiment that do not appear in the other, though the comparison graph makes this clearer by showing these vertices are completely one color or the other.